# Frequency and Time Dependence of Linear Polarization in Turbulent Jets of Blazars

**Alan P. Marscher [1,\*] and Svetlana G. Jorstad [1,2]**

[1] Institute for Astrophysical Research, Boston University, Boston, MA 02215, USA; jorstad@bu.edu

[2] Astronomical Institute, Saint Petersburg State University, Universitetsky Prospekt, 28, Petrodvorets, 198504 St. Petersburg, Russia

[\*] Correspondence: marscher@bu.edu; Tel.: +1-617-353-5029

**Abstract:** Time-variable polarization is an extremely valuable observational tool to probe the dynamical physical conditions of blazar jets. Since 2008, we have been monitoring the flux and linear polarization of a sample of gamma-ray bright blazars at optical frequencies. Some of the observations were performed on nightly or intra-night time-scales in four optical bands, providing information on the frequency and time dependence of the polarization. The observed behavior is similar to that found in simulations of turbulent plasma in a relativistic jet that contains a standing shock and/or a helical background magnetic field. Similar simulations predict the characteristics of X-ray synchrotron polarization of blazars that will be measured in the future by the Imaging X-ray Polarimetry Explorer (IXPE).

## 1. Introduction

Observations of linear polarization (LP) of blazars provide a probe of the underlying magnetic field structure that is thought to play an important role in the dynamics of the relativistic jet flows, as well as particle acceleration, in these objects. The current paradigm for the production of jets involves magnetic fields that are twisted by differential rotation of the ionized gas in orbit about the black hole (e.g., [1–3]). The magnetic field should therefore be strong and helical from the base of the jet at least out to the end of the acceleration/collimation zone (ACZ; [4–6]), beyond which turbulence might destroy—or be superposed on–the helical ordering. Such turbulence may result, e.g., from current-driven instabilities (e.g., [7]) or interactions with the external medium. Since the jets of blazars are observed to accelerate out to parsec scales [8], the magnetic launching model predicts that the helical field geometry should persist out to these scales, and that the greatest Doppler beaming of the radiation can occur parsecs from the black hole.

Most blazar jets contain a stationary compact "core" near the upstream end, plus bright "knots", some of which are essentially stationary, while others move down the jet with superluminal apparent velocities (e.g., [9]). The core seen on microwave very long baseline interferometric (VLBI) images can be modeled as a standing, roughly cone-shaped "recollimation" shock [10] or the site where the jet becomes opaque to synchrotron self-absorption [11], depending on frequency. The knots can be explained as propagating "internal" shocks [12] or simply as "plasmoids" of enhanced magnetic field and/or density of relativistic electrons. The optical synchrotron emission in blazars can arise from the core or knots, or possibly from the ACZ upstream of the core, where the predicted helical magnetic field should dominate over any turbulent component.

Each of the structures mentioned above has a characteristic predicted LP pattern that can be compared with observations:

1.  A plasmoid would likely have a slowly changing mean magnetic field direction, which might be altered in the observer's frame if the plasmoid accelerates, causing a modest (<180°) rotation of the position angle $\chi$ of LP [12]. The LP is expected to show only slight, if any, frequency dependence across the optical bands.

2.  A moving shock amplifies the magnetic field perpendicular to the shock normal (i.e., parallel to the shock front). If the shock normal is aligned with the jet axis, and if the magnetic field ahead of the shock is highly tangled, $\chi$ will be observed to be parallel to the jet axis unless the shock is viewed exactly face-on (in which case the degree of LP could be nearly zero if the pre-shock magnetic field is disordered). If the shock normal is oblique to the jet axis, $\chi$ can be oblique as well [13]. Acceleration of electrons (which includes any positrons that are present) at the shock front, followed by radiative energy losses that are more severe for higher post-shock energies, causes stratification such that the higher-frequency emission occurs over a smaller layer behind the shock front. This steepens the spectrum and increases both the mean LP and the level of variability [14]. It is possible that the ordering of the magnetic field transverse to the shock normal decays with distance from the shock front, in which case the LP can become weaker and change in position angle [15]. The gradient of maximum electron energy with distance from the shock front then causes a decrease in LP as the transverse field decays that is more pronounced at higher frequencies. Subsequently, the degree of LP reaches a minimum of zero followed by a switch of $\chi$ to 90° from the jet direction, both of which occur first at higher frequencies.

3.  A helical magnetic field tends to produce LP with $\chi$ oriented either parallel or perpendicular to the jet axis [16]. An off-axis emission feature (e.g., a slow magnetosonic shock) can cause rotation of $\chi$ by >180° as it spirals down the helical field pattern, whose field lines propagate at nearly the speed of light [17]. Unless the twist of the helical field reverses, such rotations should always be in the same sense (e.g., clockwise).

4.  Plasma with a randomly tangled magnetic field should have a low mean degree of LP, $\langle P \rangle \approx P_{maz}(N_{cells})^{-1/2}$, where $N_{cells}$ is the number of magnetic cells, each with a random magnetic field direction [18] and $P_{maz}$, which depends weakly on the slope of the synchrotron spectrum, is typically in the range of 68–75% [19]. The observed value of $P$ is predicted to fluctuate over time with a standard deviation $\sigma(P) \approx \langle P \rangle /2$ [19]. The value of $\chi$ should vary across the full range, often mimicking systematic rotations that can exceed 180°, but occurring in random directions (i.e., either clockwise or counter-clockwise).

5.  A standing conical shock compresses the magnetic field behind the shock front to form a distinctive LP pattern if the emission is resolved with VLBI. At a typical viewing angle relative to the jet axis, the pattern can be described roughly as radial [18,20,21].

6.  Dong et al. [22] carried out simulations of emission resulting from kink instabilities, finding that the degree of LP is anti-correlated with the flux density during flares. Independent simulations of kink instabilities by [23]) feature time-dependent differences between the optical and X-ray LP parameters of blazars whose X-ray emission is mainly synchrotron radiation.

Testing of these predictions requires monitoring of the flux density and LP of blazars at multiple frequencies with intensive time coverage. Here we present results of such monitoring observations along with some simulations generated with the numerical Turbulent Extreme Multi-Zone (TEMZ) model [19,24]. We compare the empirical trends with the predictions of TEMZ and other models.

## 2. Methods: Observations

The authors carried out a comprehensive multi-waveband monitoring program of 37 $\gamma$-ray bright blazars from 2008 to mid-2020. (A modified version of the program, with changes in the source list and VLBI monitoring strategy, continues.) The sample was selected based on $\gamma$-ray flux, optical flux, and declination (accessible with northern-hemisphere telescopes). The study reported here focuses on eleven of these objects that we observed at multiple optical frequencies either on several consecutive nights or multiple times during a single night.

### 2.1. Multi-Frequency Optical Observations of Linear Polarization

The optical LP data discussed here were obtained by us with the 1.83-m diameter Perkins Telescope, which is located in Flagstaff, Arizona, USA, and was operated by Lowell Observatory through 2018, after which Boston University assumed ownership and operation. The flux density and polarization was measured within the Johnson I, R, V, and B bands with the PRISM camera (www.bu.edu/prism), which includes a polarimeter with a half-wave plate that is rotated through four position angles to measure the Stokes $Q$ and $U$ parameters. The instrumental polarization was determined by observations of unpolarized stars, while the LP position angle was calibrated with polarized stars in the same field as the blazar. Corrections for optical extinction and, for Mkn421 and BL Lac, host galaxy starlight, have been made as described by [25].

On most nights, polarization measurements were carried out only at R band. During nights when all four filters were used, the observations cycled through the bands, so that a measurement at one band was offset by 4–40 min from that of another band. Because of this, frequency and time dependence can be entangled if changes in polarization occur on time-scales shorter than one hour during periods of intra-night variability.

### 2.2. VLBI Images with the Very Long Baseline Array

Throughout the ~12 years of optical observations reported here, we imaged roughly once per month the millimeter-wave structure of the jets of the blazars on scales as small as ~0.1 milli-arcseconds (mas) with the Very Long Baseline Array (VLBA) at a frequency of 43 GHz (VLBA-BU-BLAZAR program) (see Supplementary Materials). The various steps of the data processing are described by [26]. This results in images in both total and linearly-polarized intensity with an angular resolution that is typically ~0.15 mas in the east-west direction and 0.2–0.4 mas (depending on declination) in the north-south direction. Here we use the images to compare the position angle of polarization $\chi$ seen in the images with the optical values.

### 2.3. Gamma-ray and X-ray Fluxes

In this study, we present X-ray and $\gamma$-ray light curves in order to assess the level of non-thermal activity during the periods of closely-sampled optical polarization measurements. We have derived $\gamma$-ray fluxes at photon energies from 0.1 to 200 GeV from publicly available data measured by the Large Area Telescope of the *Fermi* Gamma-ray Space Telescope. We processed the data in the standard manner using software, the latest $\gamma$-ray background models, and analysis "threads" supplied by the Fermi Science Support Center (fermi.gsfc.nasa.gov/ssc/data/). The flux was calculated with a maximum-likelihood algorithm that assigns detected photons to the various sources in the field from the LAT 10-year source catalog (4FGL-DR2).

We have calculated X-ray fluxes from data obtained by the X-ray Telescope (XRT) of the *Neil Gehrels Swift Telescope*, which includes observations proposed by us and publicly available data from other programs. The data processing employed the latest versions of the HEAsoft data analysis software and CALDB calibration data base. Photons were extracted from a circular region of radius 70″ surrounding the source position, and the background was determined from an annulus of inner radius 88″ and outer radius 118″ centered on the source. The background-subtracted photon counts per energy channel data were fit with the program XSPEC in terms of a power-law spectrum with photoelectric absorption based on the Galactic foreground neutral hydrogen column density.

*2.4. Optical Flux Density Measurements with the Liverpool Telescope*

We add to the long-term optical R-band light curves fluxes measured with the 2.0-m diameter Liverpool Telescope at the Observatorio del Roque de Los Muchachos on the Canary Island of La Palma, Spain. I. M. McHardy was the principal investigator of the observing proposals.

## 3. Results of Observations

Figure 1a–k present the long-term γ-ray, X-ray, and optical R-band light curves of eleven blazars—those for which we have collected enough high-quality multi-frequency optical polarization data to allow meaningful comparison with theoretical expectations–as well as R-band degree $P_{opt}$ and position angle $\chi_{opt}$ versus time of linear polarization, the short-term variations of $P_{opt}$ and $\chi_{opt}$ during a period of intensive (either nightly or intra-night) multi-frequency optical observations, and the VLBA image at the nearest epoch to the intensive optical monitoring. The latter were scheduled in advance, and the object selected for the intensive monitoring was one that was found to have the highest flux state relative to its average. However, this close monitoring only sometimes occurred near the peak of a long-term outburst.

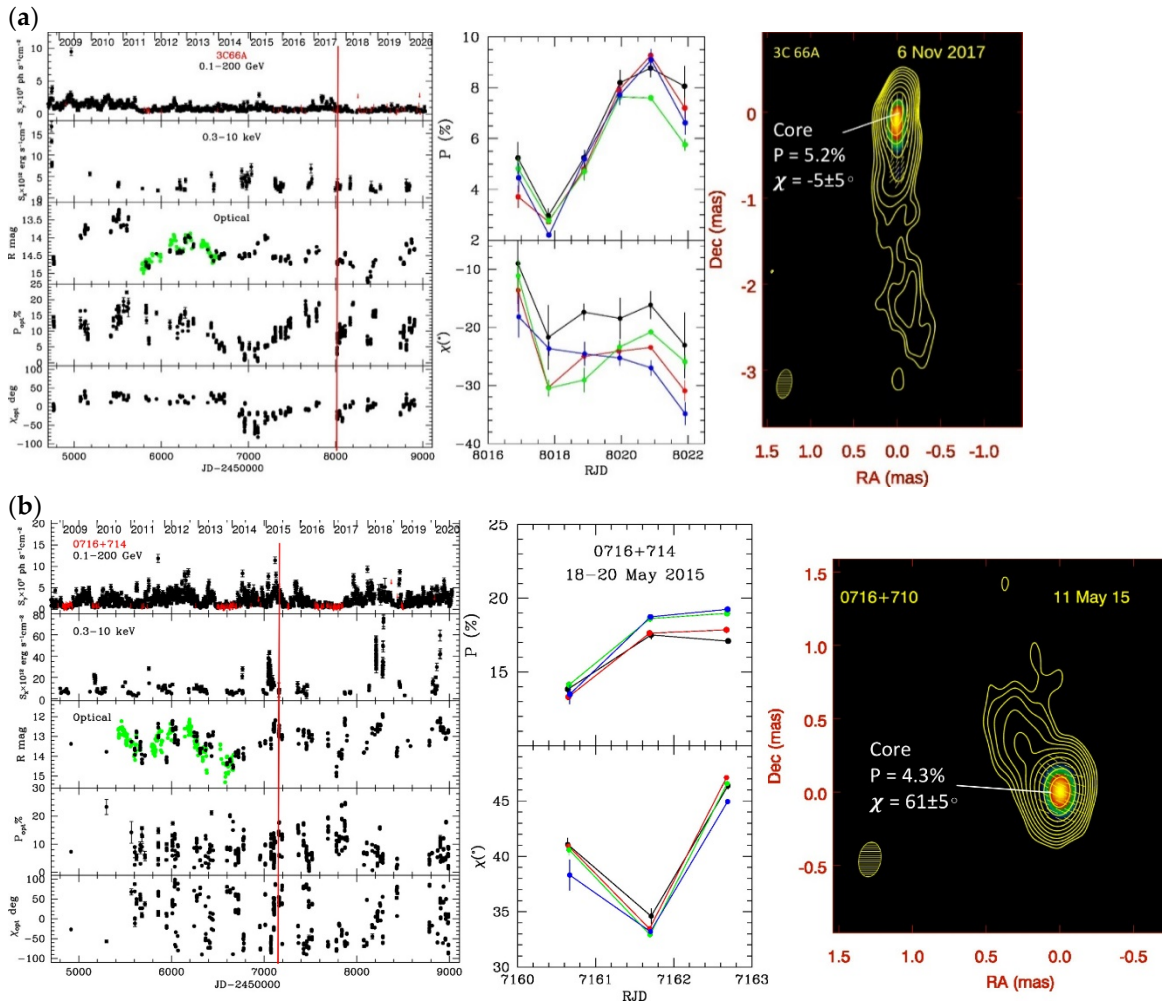

(c)

(d)

(e)

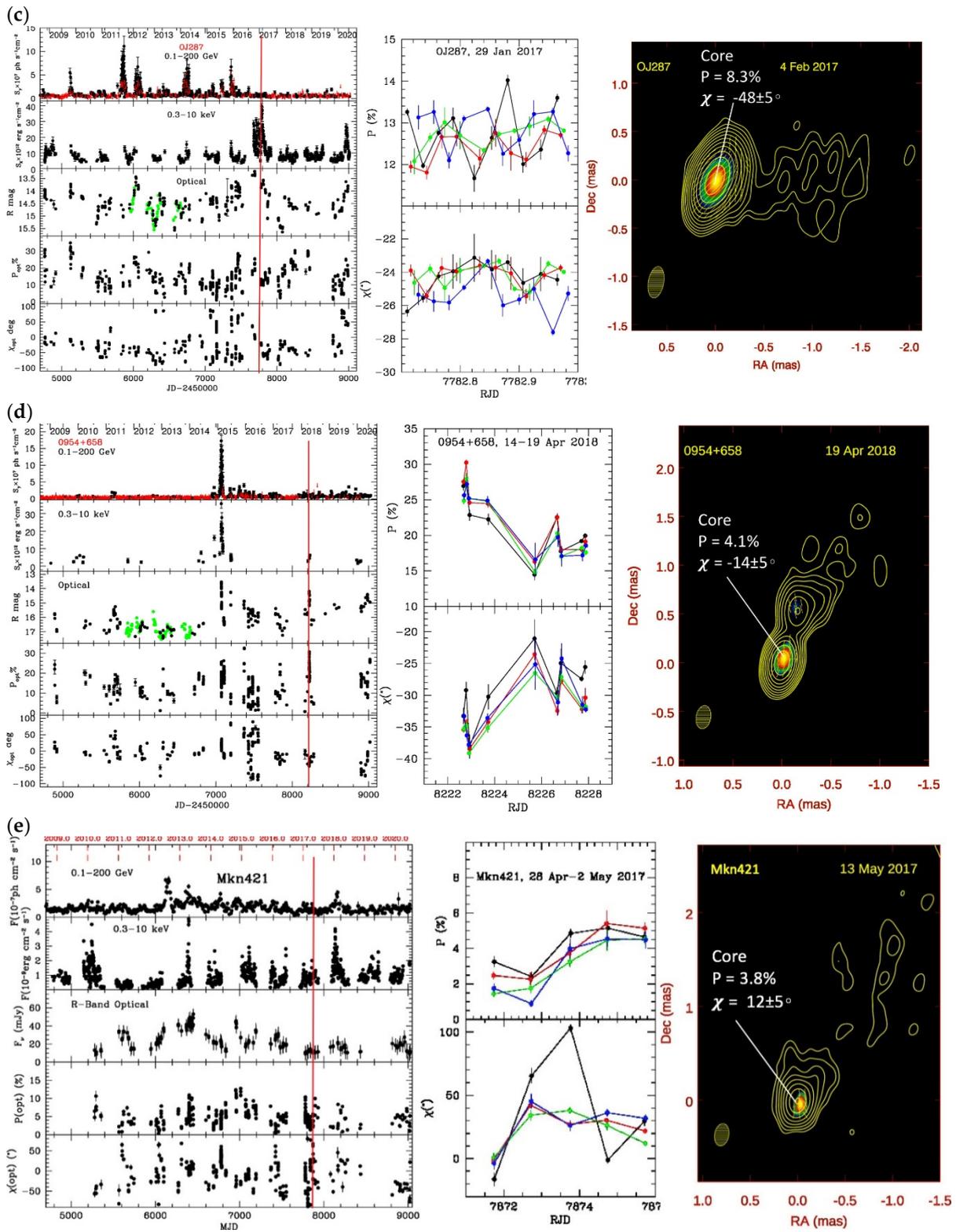

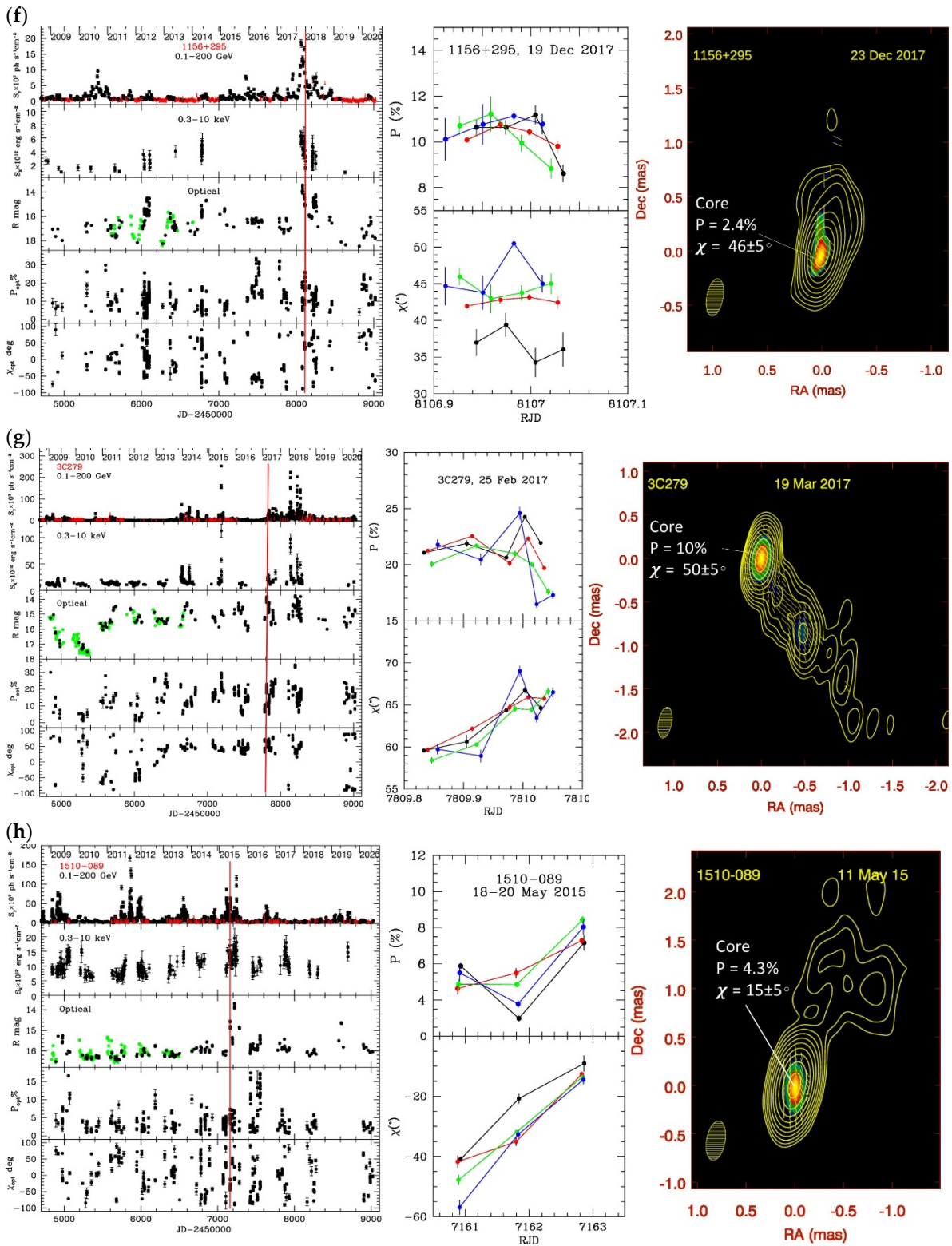

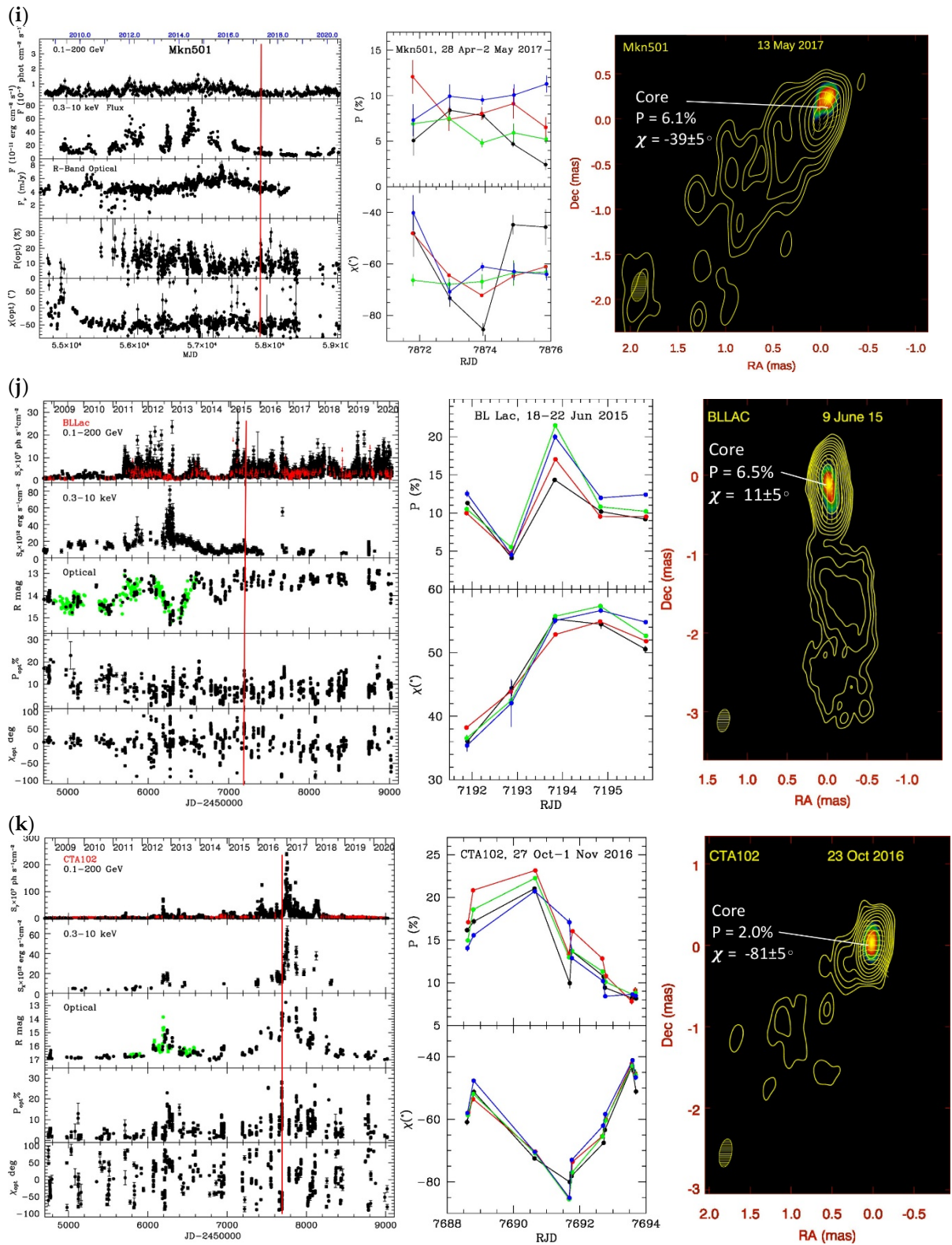

**Figure 1.** Observations of the blazars (**a**) 3C 66A, (**b**) 0716 + 714, (**c**) OJ287, (**d**) 0954 + 658, (**e**) Mkn421, (**f**) 1156 + 195, (**g**) 3C 279, (**h**) PKS 1510−089, (**i**) Mkn501, (**j**) BL Lacertae, and (**k**) CTA102. (**Left**) Long-term γ-ray, X-ray, and R-band optical light curves and optical polarization vs. time (red points in the 0.1–200 GeV light curves are upper limits). (Middle) optical polarization vs. time during a period of intensive optical monitoring (marked by the red vertical line on the light-curve panel), with black representing I band, red R band, green V band, and blue B band. (**Right**) VLBA image at the epoch of the VLBA-BU-BLAZAR program that is closest to the time of the intensive optical monitoring. Optical monitoring data are from the Perkins Telescope, except for green flux density points, which are measurements obtained with the Liverpool Telescope. For any data points without visible error bars, the errors are smaller than the symbols. For the VLBA images,

the elliptical Gaussian restoring beam (which approximates the FWHM resolution in different directions) is displayed in the lower left corner. Contours represent total intensity $I$ (in factors of 2, with the highest contour at 64% of the peak intensity) and color coding corresponds to polarized intensity $IP$; the peaks of $I$, $IP$ in Jy/beam are (**a**) 0.36, 0.019, (**b**) 1.68, 0.072, (**c**) 6.49, 0.54, (**d**) 0.44, 0.018, (**e**) 0.24, 0.091, (**f**) 1.61, 0.039, (**g**) 5.38, 0.54, (**h**) 2.48, 0.11, (**i**) 0.27, 0.016, (**j**) 1.89, 0.12, and (**k**) 2.70, 0.54.

The diversity in the behavior of the different blazars, four of which are quasars and seven of which are classified as BL Lac objects, is striking. The degree of optical polarization $P_{opt}$ is highly variable in all 11 objects, but in six of the blazars (3C 66A, OJ287, 0954 + 658, 3C 279, Mkn501, and BL Lac) $\chi_{opt}$ is often stable to within $\sim \pm 20°$ of the jet direction (as measured on a scale of <1 mas) for years, but also highly variable some of the time. In the other five cases (0716 + 714, Mkn421, 1156 + 195, PKS 1510–089, and CTA102), there is no discernable pattern in the long-term behavior of the optical polarization. The degree of optical polarization is usually higher than that of the 43 GHz core. For the objects without long-term trends in the optical polarization, this implies that the optically emitting region involves fewer turbulent cells than in the radio core (see item 4 in Section 1). In four out of five of these no-pattern blazars, $\chi_{opt}$ is within 20° of $\chi_{core}$ in the nearest-epoch 43 GHz VLBA image (PKS 1510–089 is the exception). It may be common, then, for the optical synchrotron emission in blazars to arise from a portion of the core region [27]). It is also possible that an ordered component of the magnetic field in the core is present upstream of the core, e.g., a helical field associated with the acceleration and collimation zone of the innermost jet, as inferred previously in the cases of BL Lac and PKS 1510–089 [28,29].

Significant night-to-night and intra-night variability of the polarization occurs at various levels for all of the 11 blazars. These short-term variations include significant frequency dependence, only some of which follows the trends expected from time variability coupled with the offsets in time of the measurements in the different bands.

## 4. Discussion

### 4.1. Polarization of Synchrotron Emission from a Turbulent Relativistic Jet

The observed degree of linear polarization, which is highly variable and always much less than the uniform-field case of 68–75% for synchrotron radiation, requires a strong disordered component of the magnetic field. On the other hand, the frequent alignment (to within ~20°) of $\chi_{opt}$ with the jet direction and $\chi_{core}$ implies that there is usually an ordered component as well. (Note that $\chi_{core}$ is similar to the jet direction except possibly for 1156 + 295, whose jet direction is northward but otherwise ambiguous. The inner jet of OJ287 is oriented toward the northwest, which is not apparent in the image in Figure 1c [30].) Alignment of the polarization with the jet direction can be attained if the ordered component is aligned perpendicular to the jet axis, after correcting for relativistic aberration (e.g., [16]). Such an alignment could correspond to a helical magnetic field (e.g., [16]), compression by a moving [31] or standing [20] shock, or by some other process.

Based on these considerations, we interpret our observational results in terms of a superposition of ordered and disordered magnetic fields. We do so by employing the numerical TEMZ model [19,24]. The model features thousands of turbulent cells of plasma that is flowing down a jet at a relativistic velocity. Optionally, the plasma can cross a standing shock (which is likely to be cone-shaped (e.g., [20,24])). A helical magnetic field of adjustable relative strength and geometry can be superposed on the turbulent component; see [32] for evidence indicating the presence of a helical component. The TEMZ model includes radiative energy losses of the electrons, as well as (optionally) selective acceleration of electrons: when particle acceleration occurs as the turbulent plasma crosses a shock, one expects more efficient diffusive shock acceleration of the highest-energy particles in locations where the local magnetic field is nearly parallel to the shock normal (the so-called "subluminal" particle acceleration regime), since the particle then has time to execute multiple crossings of the shock front [33,34]. This causes a volume filling factor that is lower at higher frequencies, which in turn steepens the synchrotron spectrum at optical to X-ray frequencies and the inverse Compton spectrum at γ-ray photon energies. The smaller volume at higher frequencies affects the polarization if the magnetic field has a strong turbulent component, since fewer turbulent cells are involved. As discussed in § 1, this lower value of $N_{\text{cells}}$ increases the mean degree $P$ of LP as well as the amplitude of time variations of $P$ and $\chi$.

The TEMZ code calculates the multi-frequency flux density from synchrotron radiation and inverse Compton scattering, as well as the LP of the synchrotron emission at various frequencies, as a function of both time and location within the jet. The code's spatial grid consists of up to 168 columns of cells across the jet and 112 cells along each column. Each cell has its own field direction. The turbulent magnetic field and density are implemented via a scheme similar to (although more continuous than) that of [35]. Each cell belongs to 4 nested zones of different dimensions. For each zone, the density $n_e$ and turbulent magnetic field vector $\mathbf{B}_{\text{turb}}$ are selected at random from a log-normal distribution (e.g., [36,37]) at the upstream and downstream zone boundaries. These parameters are varied smoothly in between the boundaries. A given cell's density and magnetic field vector is then the sum of the values of its zones, weighted according to the Kolmogorov spectrum (which gives higher weight to the larger zones), as found in simulations of relativistic magneto-hydrodynamic turbulence (e.g., [38]). This scheme produces the "spaghetti" magnetic field pattern characteristic of turbulence [39] for an example. In this study, we employ a version of the code that is a modification of that described in [19]. It includes a turbulent (vortical) component of the velocity of each zone [40,41], which is relativistically added to the systemic velocity vector of the jet flow. The turbulent velocity is determined in a similar manner as the turbulent magnetic field, with the Kolmogorov spectrum applying to the 4-velocity $\mathbf{v}_{\text{turb}}$ [42,43]. The physically motivated specification of a log-normal distribution of $n_e$ and $B_{\text{turb}}^2$ of the largest zones and a Kolmogorov-spectrum dependence of the quantities on zone size restrict sthe number of free parameters of the model.

Time variability of the emission from the model jet occurs as the values of $\mathbf{B}_{\text{turb}}$, $n_e$, and the maximum energy of relativistic electrons change in each cell from one time step to the next. The magnitude and direction of $\mathbf{B}_{\text{turb}}$ are both important, since the synchrotron emission coefficient is proportional to $(B \sin \phi)^{1+\alpha}$, where $\phi$ is the angle between the magnetic field and the line of sight as measured in the plasma frame, and $\alpha$ is the spectral index, with flux density $F_\nu \propto \nu^{-\alpha}$, where $\nu$ is frequency. At higher $\nu$, the number of cells with electrons of sufficient energy to radiate at that frequency decreases, i.e., $N_{\text{cells}}(\nu)$ is lower, so $\langle P \rangle$ and $\sigma(P)$ are both higher (see § 1). Fluctuations in $N_{\text{cells}}(\nu)$ caused by the random changes in $\mathbf{B}_{\text{turb}}$ and $n_e$—and therefore in the ability of individual cells to contribute to the emission at $\nu$—lead to variations in $F_\nu$, $P$, and $\chi$.

We have run the TEMZ code a number of times to sample how the time and frequency dependence of LP depends on the presence of a standing shock and/or ordered helical magnetic field component. Here we report on some trends that we have found and relate them to the data presented in Section 2. We plan a more extensive survey of parameter space in the future to explore the full range of predicted behavior and to determine whether, and how, different models that produce some similar observational signatures e.g., $\langle P \rangle$, $\sigma(P)$, $\langle \chi \rangle$, and $\sigma(\chi)$ can be distinguished from each other.

### 4.2. Results of TEMZ Simulations

We present results of simulations of some TEMZ runs that can be compared with the LP observations presented in Section 2. The selected parameters correspond roughly to a typical observed spectral energy distribution (SED) of a blazar, which naturally leads to strong variability based on the time-scales of energy losses of the radiating electrons. Here we focus on how the presence of a standing conical shock and relative levels of turbulent versus helical magnetic field components affect the behavior of the LP. For examples of light curves and SEDs from TEMZ simulations, see [19,24,44].

Figure 2a–e present (left) a scatter plot of the difference of $P$ and $\chi$ at optical B band and I band, and (right) a sample 2-day polarization vs. time curve for optical I, R, and B bands. The latter was extracted over a random 16 time-step interval (from a total of 5000 time steps per run), each of which is represented in the scatter plot. The runs all use the same angle between the jet and line of sight (3°), angle of standing conical shock front to the jet axis (8°), pre-shock bulk Lorentz factor (6), and pitch angle of helical field $\psi$ (90°, so the ordered component of the field is toroidal, as approximately expected if rotation of the flow winds the field tightly in the outer ACZ). A maximum turbulent velocity of $\sqrt{3/4}c$ (Lorentz factor of 2) is set for these simulations; that level is rarely reached when the velocity in each zone is selected randomly and velocity vectors from different zones are added. When the magnetic field is 99% helical, the variations with both time and optical frequency are very slight. The variations are stronger as the helical field is weakened to an intermediate level of 40% of the total field. At lower levels of helical field, however, the variations are only marginally stronger. This is because of the presence of the shock: turbulent plasma that is compressed by such a shock affects the LP in a similar manner as the helical field for the selected viewing angle and Lorentz factor, favoring values of $\chi$ that are more aligned with the jet direction [16,20]. The scatter plot of $P(I) - P(B)$ vs. $P(I)$ indicates that the mean polarization is higher at higher optical frequencies, an effect that is stronger when $P(I)$ is higher. This is caused by the smaller number of cells with electrons of high enough energy to radiate at higher frequencies, as discussed above. The simulated frequency dependence of $P_{opt}$ is weak when the helical field is stronger than ~50%, while the frequency dependence of $\chi_{opt}$ is weak until the helical field is less than ~30% of the total.

**(a)**

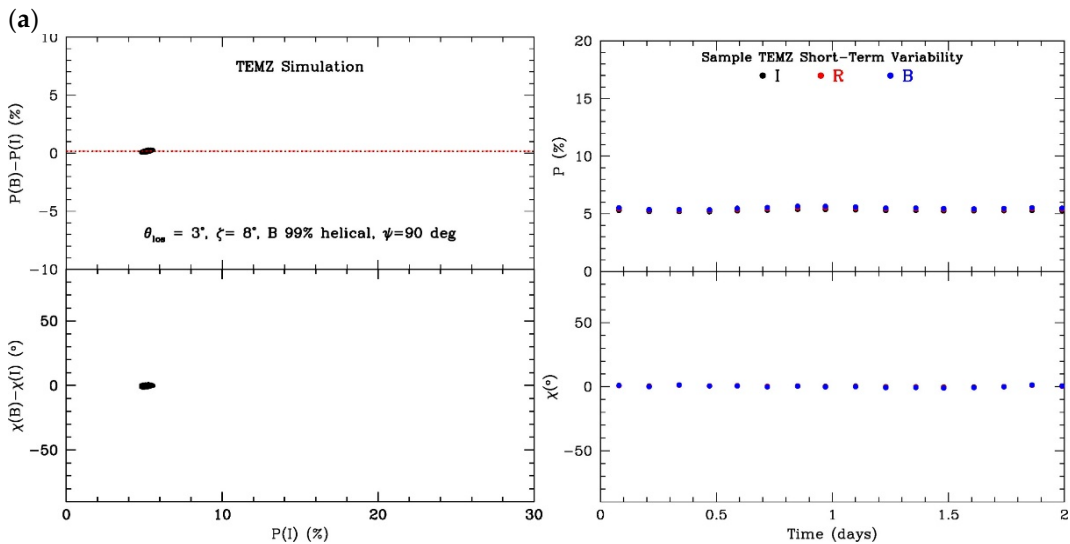

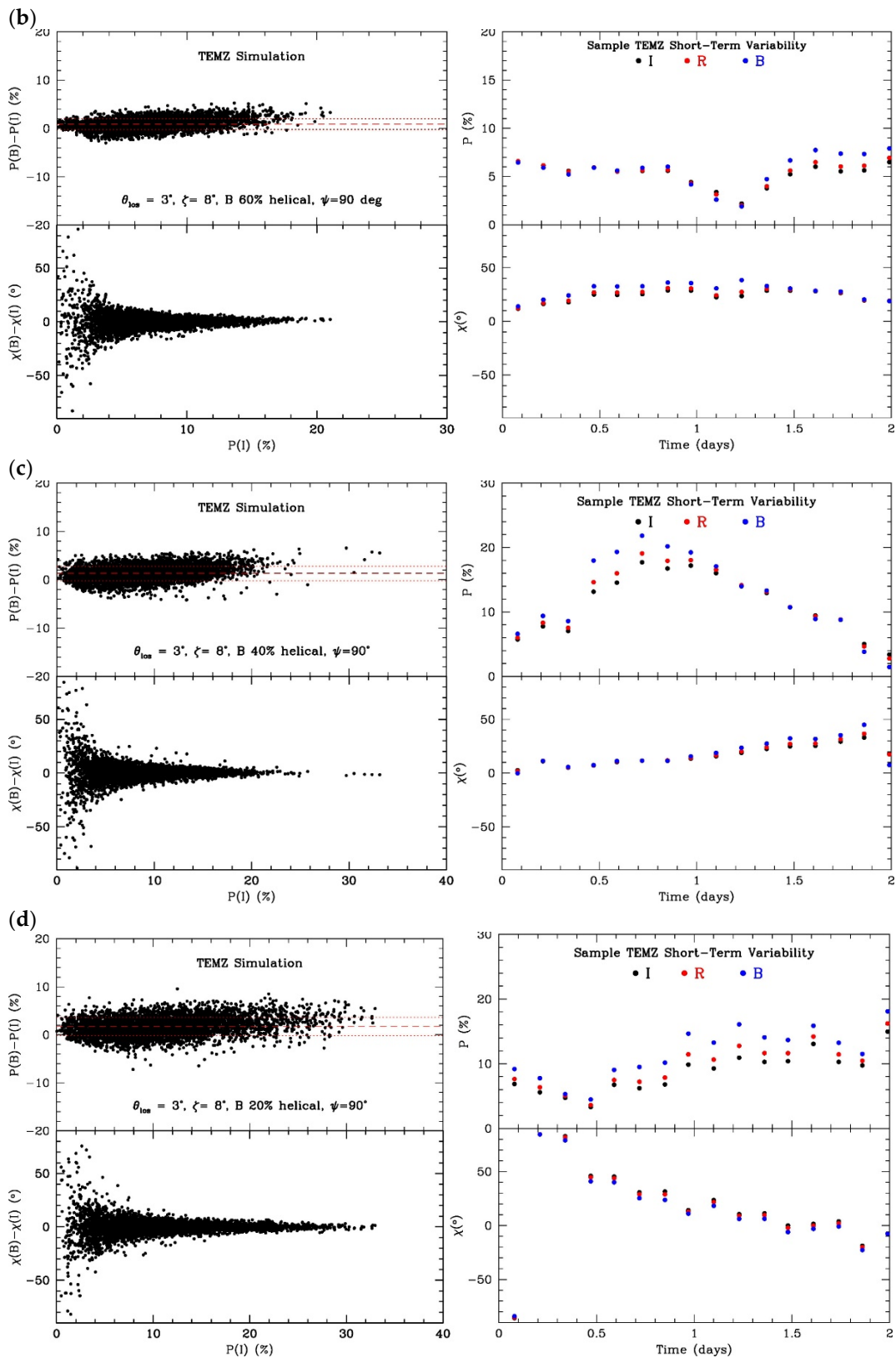

**(e)**

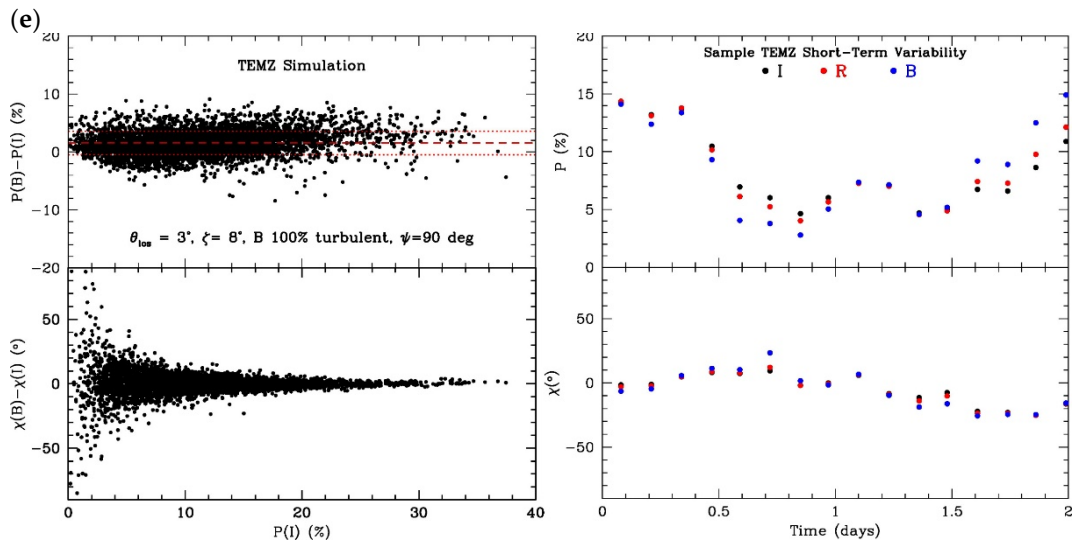

**Figure 2.** (**Left**) frequency and (**right**) short-term time dependence of the optical linear polarization of a turbulent jet with a standing conical shock representing the "core." The ratio of ordered helical to total (helical + turbulent) magnetic field decreases from (a) 99% to (e) 0%. The dashed red horizontal line in the upper left panels indicate the mean value of [$P(\mathrm{I})$–$P(\mathrm{B})$], while the dotted lines demark $\pm 1\sigma$.

Figure 3a–c display similar plots for a jet with no shock for three different ratios of helical to total magnetic field. The emission zone is cylindrical, and electrons are accelerated with a power-law energy distribution at the upstream boundary of the zone (perhaps from many magnetic reconnections), beyond which they lose energy as they radiate. Without the shock compression, the polarization is more strongly variable, as can be seen in the right-side panels of Figure 3a–c.

**(a)**

**(b)**

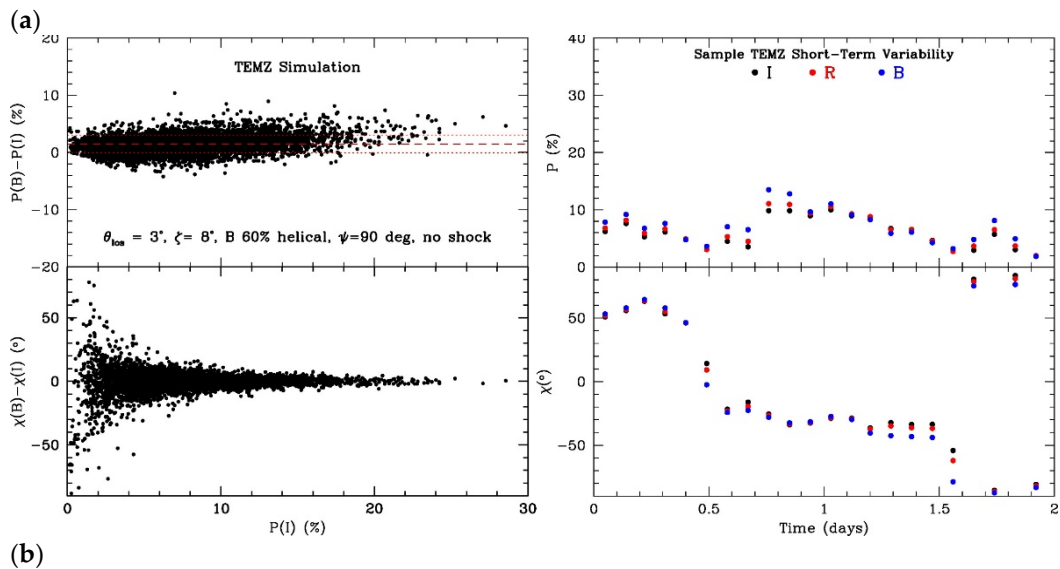

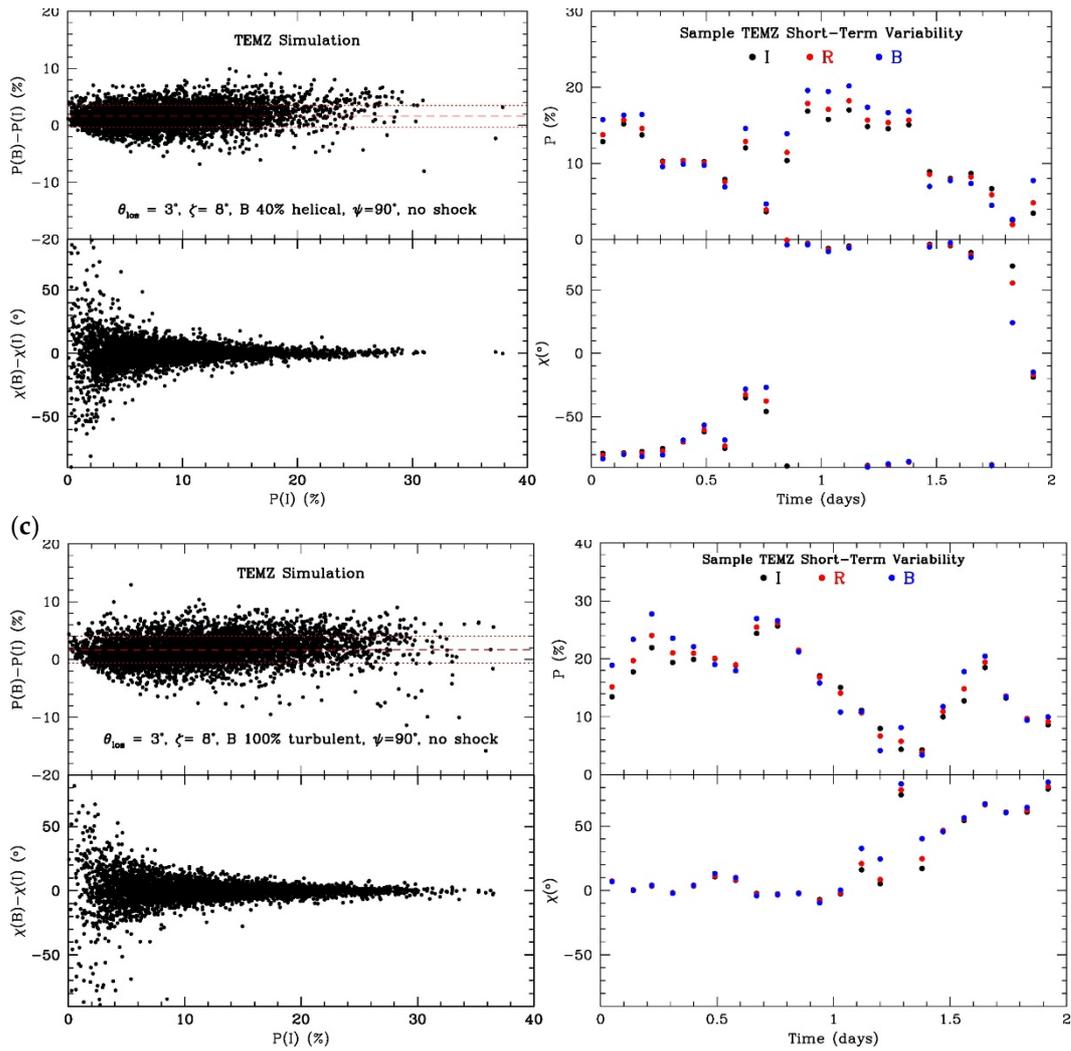

**Figure 3.** Similar to Figure 2 for the case of a turbulent jet without a shock and for only three ratios of helical to total magnetic field: (**a**) 60%, (**b**) 40%, and (**c**) 0% (100% turbulent field).

Figure 4 reveals a criterion for differentiating between the different levels of helical field and the standing shock vs. no-shock case, based on the long-term dependence of $\chi_{opt}$ on time. The presence of either a standing conical shock or a helical magnetic field favors values of $\chi_{opt}$ that are parallel to the jet axis. For the viewing angle of 3° adopted in the simulations, the effect of the helical field is weaker than it would be if the viewing angle were closer to the inverse of the Lorentz factor (1/6 radians ~10°) [16]. As expected, the case of 100% turbulent magnetic field with no shock (blue color in the right-side panel) yields a random distribution of $\chi_{opt}$ values.

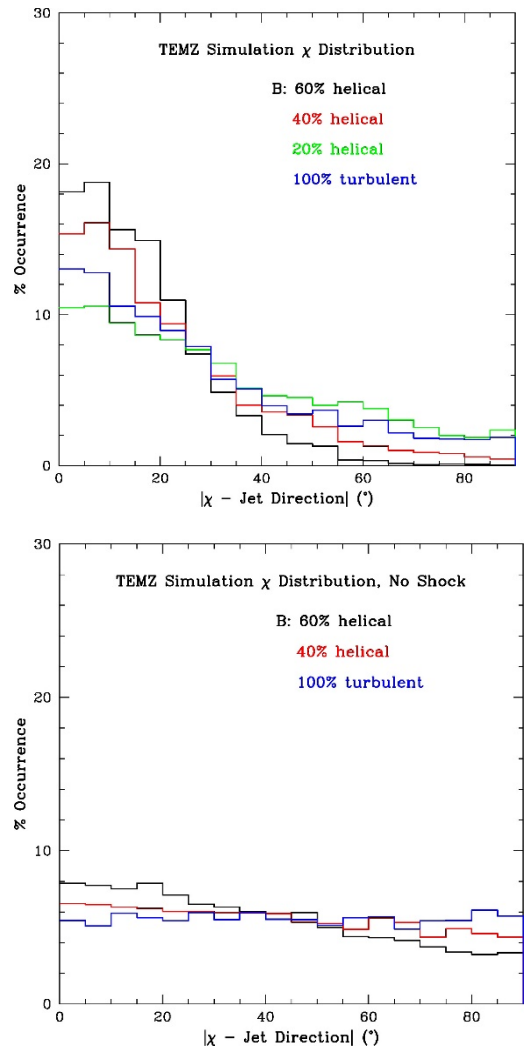

**Figure 4.** Histograms of the distribution of χ(I) relative to the jet direction from the same TEMZ simulations as in Figures 2 and 3. Left panel: turbulent jet with a standing conical shock. Right panel: turbulent jet without a shock.

### 4.3. Comparison of Simulations with Data

Although we have not yet amassed enough multi-frequency optical LP monitoring data to compare statistically with the simulations, inspection of Figures 1–4 allows us to draw some general conclusions regarding the ability of different models to explain the observational behavior of the blazars in our sample. It is obvious that the case of 99% helical field (Figure 2a) is incompatible with the level of variability of LP of the blazars. The mild variability and frequency dependence of the standing shock plus 60% helical field case (Figure 2b) are somewhat weaker than observed. However, the no-shock, 60% helical field case (Figure 3a) produces a sufficient level of variability. Without a shock, the essentially random values of $\chi_{opt}$ observed in 0716 + 714, Mkn421, 1156 + 195, PKS 1510−089, and CTA102 (Figure 1a,e,f,h,k) cannot accommodate a helical field stronger than ~40% unless the viewing angle is ~0, but in that case $P_{opt}$ would be close to zero, much lower than observed. For the other six blazars, both a standing shock and/or a helical field up to ~60% are compatible with the data.

## 5. Predictions of Polarization of X-ray Synchrotron Emission from Blazars

The Imaging X-ray Polarimetry Explorer (IXPE), a joint NASA-Italian Space Agency mission, is scheduled for launch in late 2021. It is designed to detect X-ray (2–8 keV) LP, with a sensitivity expected to be sufficient to measure $P_x$ and $\chi_x$ of a number of blazars with high X-ray fluxes. The TEMZ code simulates the X-ray LP as a function of time when the magnetic field strength and highest electron energies are sufficient to produce a high flux of X-ray synchrotron radiation. Among the blazars studied here, Mkn421 and Mkn501, which are very bright X-ray sources (see Figure 1e,i), are prime candidates for IXPE observations. We have run the TEMZ code with parameters that roughly reproduce the SED of Mkn421. Figure 5 compares the simulated X-ray and optical LP. The variability is somewhat stronger at X-ray energies, and the degree of X-ray polarization is almost always higher at X-ray energies than at optical frequencies. This agrees with the conclusion of [45], who considered a turbulent plasma without a shock or helical field. The short time-scale of variability of LP (20 min) in the simulations implies that exposure times of hours could result in LP-vector averaging that will lead to a lower measurement of $P_x$ than actually occurs. We point out, on the other hand, that Mkn421 is less distant than the other potential IXPE blazar targets. As such, it has a lower luminosity and is physically smaller–and therefore variable on a shorter time-scale–than its higher-redshift, more luminous counterparts. Its X-ray flux is also high enough to determine whether the X-ray LP becomes stronger as one shortens the integration time of the measurement.

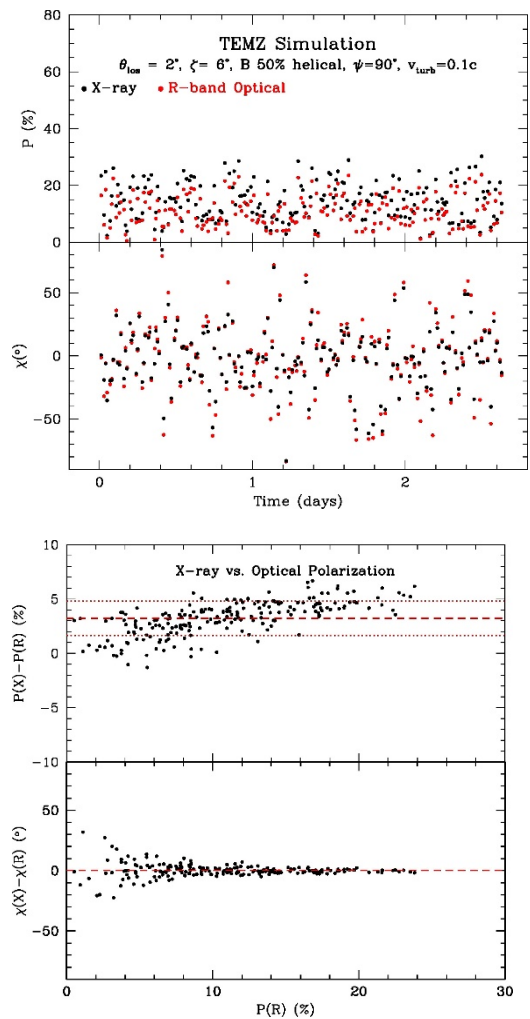

**Figure 5.** Simulated short time-scale variations of X-ray and optical linear polarization for parameters that roughly fit the SED and variability characteristics of Mkn421. The dashed red horizontal line in the upper right panel signifies the mean value of $[P(X) - P(R)]$, while the dotted lines demark $\pm 1\sigma$. The dashed line in the lower right panel demarks $\chi(X) - \chi(R) = 0$.

## 6. Conclusions

Our observations of short time-scale LP variability at different optical frequencies demonstrate that substantial intra-night and night-to-night, frequency-dependent fluctuations of $P_{opt}$ and $\chi_{opt}$ are common. Simulations with the TEMZ code produce results that resemble the observed LP behavior of our sample of eleven blazars. Successful comparison between the simulated and empirical time and frequency dependence of LP requires a substantial disordered component of the magnetic field, as expected if the jet plasma is turbulent. We thus support, with a larger sample, the conclusions of [46], who carried out similar intensive optical LP observations of 0716 + 714, and [47], who similarly observed BL Lac. On the other hand, our study also allows—or in many cases requires—an ordered component that partially aligns the field with the jet direction.

In sources such as Mkn421 where electrons are accelerated to energies high enough to produce synchrotron X-ray emission, the X-ray LP should nearly always be somewhat stronger than that at optical frequencies. X-ray LP in blazars with such emission is therefore expected to be detectable with IXPE. The X-ray polarization vector should, however, fluctuate on shorter time-scales than at optical frequencies. In some blazars this could require short integration times in order to avoid vector averaging that would significantly decrease the measured degree of LP below the actual value.

We plan to continue our study by carrying out a statistical analysis of the long-term LP observations of our full sample at optical and other frequencies. We will accompany this with an expanded exploration of the simulated behavior of the time and frequency dependence of LP with the TEMZ model. The goal is to provide a framework within which the complex emission properties of blazars can be characterized and perhaps understood.

**Supplementary Materials:** VLBA data and images with polarization information at many epochs, as well as multi-waveband light curves, of 37 γ-ray bright blazars and radio galaxies, are publicly available online at http://www.bu.edu/blazars/BEAM-ME.html. Multi-color optical polarization data are listed online at www.bu.edu/blazars/mobpol.html.

**Author Contributions:** Both authors participated in all aspects of the article. Both authors have read and agreed to the published version of the manuscript.

**Funding:** The research described here was funded by NASA through Fermi Guest Investigator grants 80NSSC20K1566 and 80NSSC20K1567, and by National Science Foundation grant AST-1615796.

**Data Availability Statement:** Data are publicly available online at http://www.bu.edu/blazars/BEAM-ME.html.

**Acknowledgments:** This study made use of data obtained at the Perkins Telescope Observatory, which is owned and operated by Boston University. The Very Long Baseline Array is an instrument of the National Radio Astronomy Observatory. The National Radio Astronomy Observatory is a facility of the National Science Foundation operated by Associated Universities, Inc.

**Conflicts of Interest:** The authors declare no conflict of interest.